\documentclass[12pt]{article}
\usepackage{amsmath}
\usepackage{epsfig}

\def\am{\left(\begin{array}{c}}
\def\amm{\left(\begin{array}{cc}}
\def\ammm{\left(\begin{array}{ccc}}
\def\ammmm{\left(\begin{array}{cccc}}
\def\a{\end{array}\right)}

\def\vecs{\mathsf}
\def\vec{\mathbf}
\def\#{\mathbf}

\def\E{\epsilon}
\def\M{\mu}

\def\K{\kappa}

\def\p3{{
\, \bigcirc {\hskip -2.7mm} {\scriptstyle 3} \;\,
}}

\def\D{\nabla}

\begin{document}

\title{Waves and energy in chiral nihility}
\author{
Sergei Tretyakov, Igor Nefedov, Ari Sihvola, \\ 
Stanislav Maslovski, Constantin Simovski}
\date{Radio/Electromagnetics Laboratory, Helsinki University of
Technology\\ P.O.\ Box 3000, FIN--02015 HUT, Finland}


\maketitle

\bigskip
{\center
\large

Address for correspondence:

Sergei A.\ Tretyakov, \\ Radio Laboratory,
Helsinki University of Technology,\\ P.O. Box 3000, FIN-02015 HUT, Finland.

\medskip
Fax: +358-9-451-2152

E-mail: sergei.tretyakov@hut.fi

}

\vskip 1cm

\begin{abstract}

A model for a chiral material in which both the permittivity and
permeability are equal to zero is discussed.
 Such a material is referred by us as
a ``chiral
 nihility".
It is shown that this exotic material can be realized as a
mixture of small helical inclusions.
 Wave solutions and energy in such a medium are
analyzed. It is shown that an extraordinary wave in
 chiral nihility is
a backward wave. Wave reflection and
 refraction on a chiral
nihility interface is also considered. It is shown that a linearly
polarized wave normally incident onto this
 interface produces the
wave of ``standing phase" and the same wave
 in the case of oblique
incidence causes two refracted waves, one
 of them with an anomalous
refraction.
\end{abstract}

\vskip 2cm

\null \hfill{\sl \today}

\newpage

\section{Introduction}

Recently, a lot of attention has been paid to composite materials
which in certain frequency regions can be described by effective
permittivity and permeability having negative real parts (or, approximately,
by real negative material parameters). In this respect,
A.~Lakhtakia conceptually considered two-phase mixtures of an ordinary
isotropic material with positive material parameters and of a double negative
material, such that the effective parameters of the mixture become null.
He introduced  term ``nihility" for such medium, whose
$\E=0,\,\mu=0$ \cite{Lakh1}. He also used this notion  in a
later paper \cite{Lakh2} when discussing the concept of perfect lens
proposed by J.B.~Pendry \cite{Pendry}.
Lakhtakia's conclusion was that nihility material is not physically realizable.

In this paper, we model a regular array of small chiral (or
$\Omega$-shaped) ideally conducting particles and find out that at a
certain frequency this system behaves as an effective media with
null-valued permittivity and
 permeability. Of course, in real
physical situations there will be
 nonzero imaginary parts of the
material parameters due to
 absorption in the particles. However, no
waves can travel in nonchiral nihility media
\cite{Lakh1}, since the
 field equations reduce to the static ones.

Furthermore, in this paper, we will generalize the concept of nihility
and introduce a more general concept of
 {\it chiral nihility}
composite materials. These materials can
 be realized in a similar way
as mixture with ``ordinary'' nihility. We have to add chiral
inclusions, for example,  all of the same handedness.
 At the
frequency where the real parts of both permittivity and permeability
become zeros, the chirality parameter is nonzero, and, as we will
demonstrate by a numerical example, the imaginary parts of all the
parameters can be
 rather small compared to the chirality parameter.
In these media, waves can propagate, and the material exhibits
 some
very interesting properties. In particular,
 double refraction takes
place at an interface between free space
 and an {\it isotropic}
chiral nihility. A linearly polarized wave is
 split into two
circularly polarized once, and one if these two components
 suffers
negative refraction, as in  backward-wave or double negative
materials.

The constitutive relations for isotropic chiral media
read \cite{chibi}
\begin{eqnarray}
{\vec D} & = & \E\E_0{\vec E} - {\rm j}\K \sqrt{\E_0\M_0} \, {\vec H} \\
{\vec B} & = &  {\rm j}\K \sqrt{\E_0\M_0} \, {\vec E} + \M\M_0{\vec H}
\end{eqnarray}
Chiral nihility media are, by definition, media with the material parameters
satisfy, at a certain frequency $\omega _0$,
\begin{equation}
  \E=0, \quad \M=0, \quad \K\neq 0
\end{equation}
Thus, the material relations reduce to
\begin{eqnarray}
{\vec D} & = & - {\rm j}\K \sqrt{\E_0\M_0} \, {\vec H} \\
{\vec B} & = &  {\rm j}\K \sqrt{\E_0\M_0} \, {\vec E}
\end{eqnarray}
The different electric and magnetic units call for a renormalization
of the quantities when there is magnetoelectric coupling. Also, to
achieve a compact notation for the material response analysis, a good
technique is the six-vector notation with which the constitutive
parameters are contained in a material matrix $\vecs M$:
\begin{equation}\label{eq:M}
 \am c\eta_0\vec{D} \\ c\vec{B} \a =
\amm \E & -{\rm j}\K \\ {\rm j}\K & \M \a \am
 \vec{E} \\ \eta_0\vec{H} \a = {\vecs M}  \am \vec{E} \\ \eta_0\vec{H} \a
\end{equation}
in other words,
\begin{equation}
{\vecs d} = {\vecs M} {\vecs e}
\end{equation}
where the fields and displacements now carry the same dimensions
(V/m), and the material matrix components are
dimensionless.\footnote{The coefficients in the renormalization
are the vacuum constants $c=1/\sqrt{\E_0\M_0}$ and
$\eta_0=\sqrt{\M_0/\E_0}$.}

\section{Model of the material parameters of chiral nihility}

\begin{figure}[h]
\centering
\epsfig{file=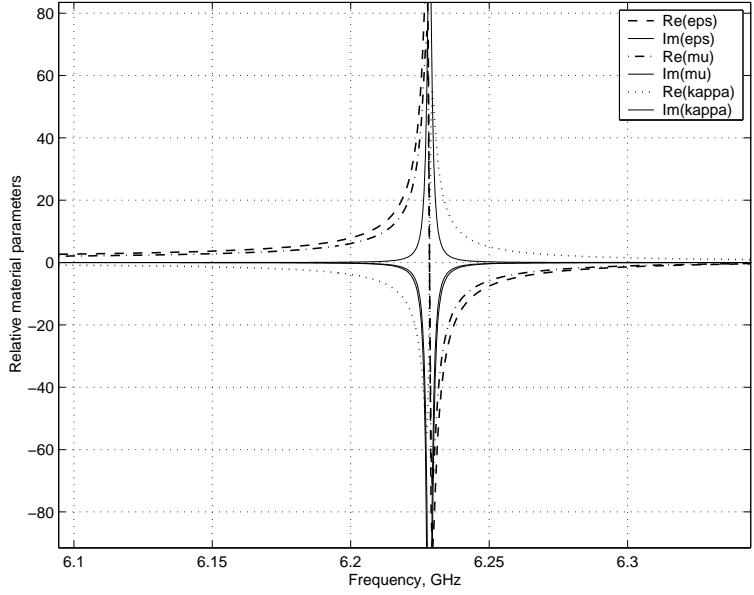, width=10cm}
\caption{Effective material parameters of a lattice of canonical helices.
See text for the details of the mixture.}
\label{wide}
\end{figure}

\begin{figure}[h]
\centering
\epsfig{file=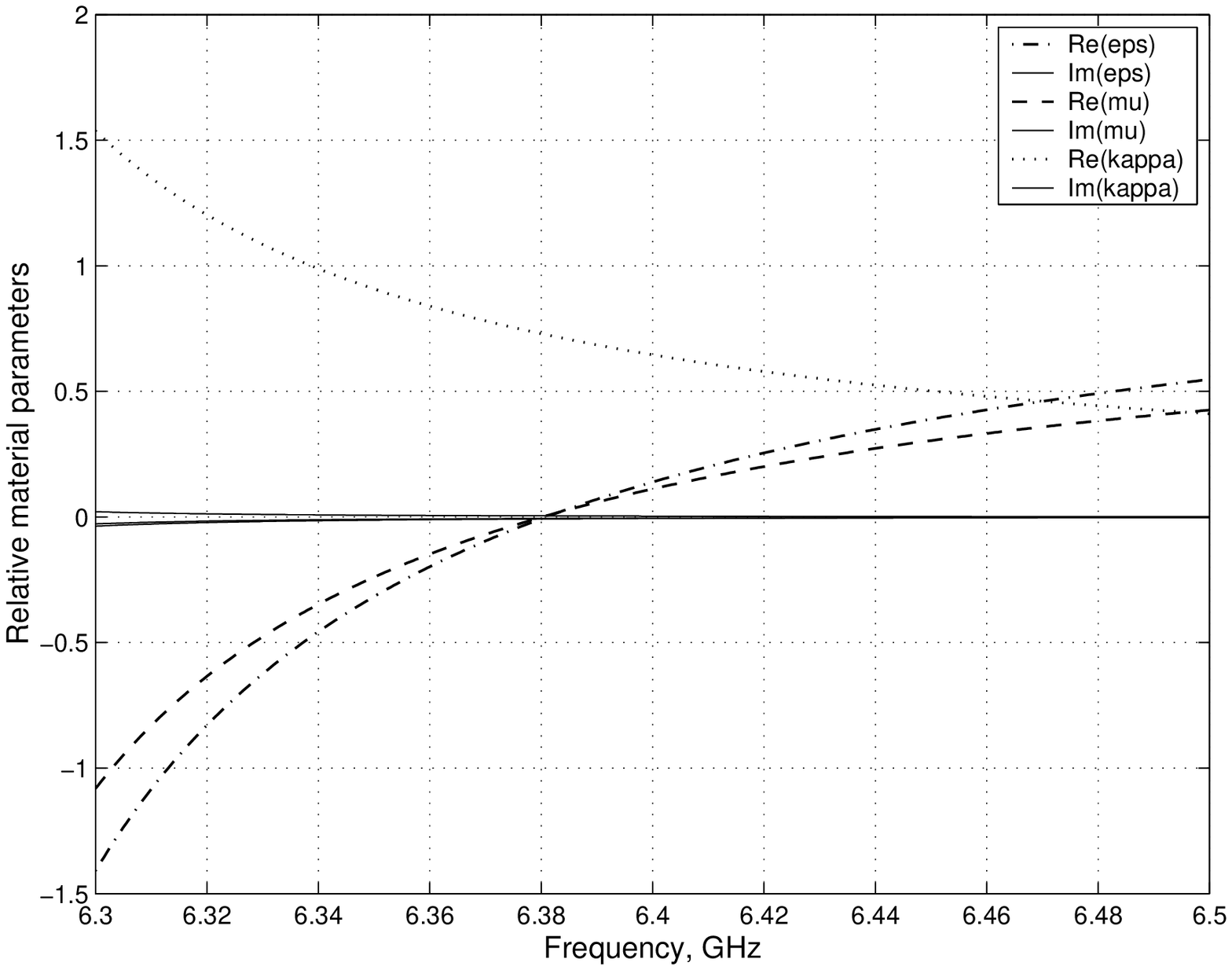, width=10cm}
\caption{The same as in Figure 1 for a narrow frequency range.}
\label{narrow}
\end{figure}

Using the antenna model of canonical chiral particles
that consist of a small loop connected to a short wire dipole antennas \cite{antenna} and the
Maxwell Garnett mixing rule, we can
 calculate the effective
parameters of an isotropic array of chiral
 particles. In these
calculations, we assume that the
 array is regular, so that the
scattering loss of the particles
 is suppressed by the interaction
field. The absorption loss
 has been taken into account by adding
appropriate real parts to the
 input impedances of the loop and wire
portions of the
 canonical chiral helix.  The geometrical parameters
have been chosen so that the
 real parts of the permittivity and
permeability become zero at the
 same frequency. The results of these
calculations are shown in Figures~\ref{wide} and \ref{narrow}.
 The
second picture is a blow-up near the frequency of zero permittivity
and
 permeability. The inclusion sizes are the following:
 the arm
length of the straight dipole $l=2.7$\; mm, the loop radius
 $a=2.45
$\; mm, the wire radius $r_0=0.25$\; mm. Wires are made of copper with
conductivity $\sigma=5.8\cdot 10^7$ S/m, and the volume fraction is
$f=0.2$. The last
 value is defined by introducing imaginary spheres
of the minimum
 radius totally enclosing the helices.

\vspace{5mm}

As is obvious from the graphs in Figures 1 and 2, at the frequency
equal to 6.38 GHz,
 both the permittivity and permeability of the
material have zero real parts.
 The chirality parameter is about 0.7
at the frequency. The imaginary parts
 are rather small compared with the chirality parameter near this
frequency and we will neglect them in the
 following analysis. In
passing, let us note that a racemic arrangement of the
 same particles
gives a {\em nihility} material with $\epsilon=0$, $\mu=0$, and
$\kappa=0$. Therefore, not only nihility medium is possible but also a
material displaying chiral nihility.

Let us next move on to study theoretical limitations for the material
parameters in complicated media.

\section{Material parameter restrictions}

The limitation for material parameters in lossless chiral media are
\cite{chibi}
\begin{equation}
\K^2 \leq \E\M \label{limit}
\end{equation}
This restriction comes from the requirement that the wave numbers
$k_\pm = k_0(\sqrt{\M\E}\pm\K)$ (for lossless media) should be
positive.
 However, why is it necessary to have both wave vectors
positive? The answer to this is not obvious.  Another way to justify
the restriction (\ref{limit}) is to consider
 the eigenvalues of the
material matrix. These are---being the
 solutions for the equation
\begin{equation}
{\vecs M} {\vecs e} = \lambda {\vecs e}
\end{equation}
for the eigenproblem---the following two values
\begin{equation}
\lambda_{1,2} =
\frac{\E+\M}{2} \pm \sqrt{ \left( \frac{\E-\M}{2}\right)^2 + \K^2 }
\end{equation}

Now, the usually accepted limitations (\ref{limit})
are seen to correspond to the requirement that
\begin{equation}
\lambda_1 \geq 0, \quad \lambda_2 \geq 0
\end{equation}
In other words, the limitation means that the material matrix $\vecs
M$ be positive definite. This, subsequently, corresponds to the
condition that the sum of electric and magnetic energy densities is
positive:
\begin{equation}
W_e + W_m \propto {\vec E}\cdot{\vec D} + {\vec H}\cdot{\vec B}
\propto {\vecs e} \cdot {\vecs M}{\vecs e}
\end{equation}
With this in mind it seems that the limitation (\ref{limit}) is rather
reasonable.  But locally we can certainly violate the
limitation. The Veselago medium, or a basic plasma medium for example,
as such violate against it. Should we have $\E\leq 0$, the property is
sufficient to cause locally negative static energy density which is
proportional to $\E|{\vec E}|^2$.
Because plasma is a well-documented real-world phenomenon, and also
a 2D analogue of Veselago media has been fabricated in laboratories
\cite{Science}, we have to
conclude that the restriction is too tight.
In fact, the nihility medium as well as Veselago medium possess
strong frequency dispersion, and the energy density should be
calculated with this in mind \cite{Landau}.  The time-averaged energy density in a dispersive chiral medium with negligible losses 
  is expressed as \cite{book}
\begin{equation}
\label{a6}
  \langle W\rangle_t=\frac{1}{4}\left[\vec E^*\cdot\frac{\partial(\omega \E)}{\partial\omega }
  \cdot\vec E+\vec H^*\cdot\frac{\partial(\omega \mu)}{\partial\omega }
  \cdot\vec H+\frac 2c{\rm Im}\left(\vec E^*\cdot\frac{\partial(\omega \K)}{\partial\omega }
  \cdot\vec H\right)\right]
\end{equation}
We will come back to the discussion of the energy density after
considering possible wave solutions in chiral nihility.

On the other hand, the requirement for the imaginary parts of the
material parameters for any medium
\begin{equation}
{\rm Im}\,\{\E\} \leq 0, \quad {\rm Im}\,\{\M\} \leq 0,
\quad {\rm Im}\,\{\K\} \leq {\rm Im} \,\{\sqrt{\E\M}\} \label{im}
\end{equation}
should hold, because the amplitude of a propagating wave should not
grow exponentially in a dissipative medium.
And this is indeed the case. According to the model discussed and
applied in the previous section, we can see that the imaginary parts
of the material
 constants satisfy (\ref{im}). Therefore also ${\rm
Im}\{k\}\rm<0$ which is
 necessary in order that the wave propagation
according to $\exp(-{\rm
 j}kz)$ is not exponentially increasing.

\section{Waves in chiral nihility materials}

Let us next analyze waves in chiral nihility media.
Taking the time dependence as $\exp({{\rm j}\omega  t)}$,
 the Maxwell equations for chiral nihility can be written as
\begin{equation}
\label{a1}
\begin{array}{l}
   \D\times\vec{E}=k_0\K\vec{E} \\
    \D\times\vec{H}=k_0\K\vec{H},
      \end{array}
\end{equation}
where $k_0=\omega /c$ is the wavevector in vacuum.
 Solutions of (\ref{a1}) are the eigenvectors of the curl operator
 and describe the circularly-polarized waves with helicity
 parameter $k_0\K$.
At first look it seems that each of equations (\ref{a1})
 determines independent ``electric" and ``magnetic" waves, like
 Langmuir waves in plasmas. However, we have to remember that the
 chiral
 nihility medium is obtained by the mixture of components 
 having
 positive and negative electric and magnetic polarizabilities
 which
 must be dispersive. Thus, we have to adopt
 that
 electric and magnetic field are not independent but are
 connected
 via the wave impedance
 \begin{equation}
 \eta=\eta_0\lim_{\E\to 0,\mu\to 0}\sqrt{\frac{\mu}{\E}}
 \label{eta}
\end{equation}
Of course in nihility the quotient of impedance has to be
 carefully defined as a limit process.  The limit $\mu/\E$ for
 $\omega \to\omega _0$ depends on the behavior
 $\E(\omega ),\mu(\omega )$ in the vicinity
 of $\omega _0$.
In our particular example (Figure~\ref{narrow}) this value is close
to the impedance of free space, because the values of the effective permittivity and
permeability are close in the vicinity of the nihility point.

 Assuming the wave propagation direction is along
 $z$-axis, solutions of equations~(\ref{a1}) can be written in form
\begin{equation}
\label{a2}
\begin{array}{ll}
 E_x=e_0\exp(\mp {\rm j}k_0\K z) &  H_x=\pm{\rm j}e_0\eta^{-1}\exp(\mp {\rm j}k_0\K
   z)\\
 E_y=\mp{\rm j}e_0\exp(\mp {\rm j}k_0\K z) &  H_y=e_0\eta^{-1}\exp(\mp {\rm j}k_0\K
   z)\\
   E_z=0   &  H_z=0
      \end{array}
\end{equation}
 where $\eta$ is the wave impedance of chiral
 nihility (\ref{eta}) and the two signs correspond to the  phase advance direction along
 the positive or
 negative direction of axis $z$. Obviously, the waves are circularly polarized with
 the opposite sense of rotation for the opposite propagation
 directions.
The waves in chiral nihility propagate with the propagation constant $\beta=\pm k_0\K$
  and the phase
velocity $v_p=\pm c/\K$,
where $c$ is the speed of light. The phase velocity $ v_p$ can be
either subluminal if $\K>1$ or superlumunal, if $\K<1$.

Let us next discuss the group velocity of plane waves in chiral nihility.
In usual  chiral media there exist two eigenwaves traveling along  the 
positive $z$-direction with the phase constants 
 $\beta_{1,2}=k_0(\sqrt{\E\mu}\pm\K)$. For the  inverse group velocity of these waves 
 we obtain
\begin{equation}
 \label{a33}
\frac{1}{v_{gr1,2}}=\frac{\partial\beta_{1,2}}{\partial\omega }={\sqrt{\epsilon\mu}\over c}+
 \frac{\omega }{c}\frac{\partial(\sqrt{\E\mu})}{\partial\omega }
 \pm\left(\frac{\K}{c}+ \frac{\omega }{c}\frac{\partial\K}{\partial\omega }\right)
\end{equation}
Usually, both these values are positive, and both eigenwaves are usual 
forward waves.
The other two eigensolutions have the oppositely directed phase vectors
($\beta_{1,2}=-k_0(\sqrt{\E\mu}\pm\K)$) and, of course, their group velocities 
are negative of that given by (\ref{a33}). At the cross-over point where
$\sqrt{\epsilon\mu}=0$ the eigensolutions become degenerate in terms of the
phase constants: we have only {\it two} waves with $\beta=\pm k_0\K$ instead of four. However, the group velocities are still given by 
(\ref{a33}) with $\sqrt{\E\mu}=0$:
\begin{equation}
 \label{a3}
 \frac{1}{v_{gr1,2}}=
 \frac{\omega }{c}\frac{\partial(\sqrt{\E\mu})}{\partial\omega }
 \pm\left(\frac{\K}{c}+ \frac{\omega }{c}\frac{\partial\K}{\partial\omega }\right)
\end{equation}
and there are still {\it four} different values for the group 
velocity of four eigenwaves (the other two solutions are the negative of 
(\ref{a3})).
If we assume that at this special point both values (\ref{a3}) are positive, 
as is generally the case in chiral media, then the conclusion is that 
one of the two eigenwaves with positive group velocity has a negative
phase velocity, that is, one of the eigenwaves is a backward wave.

Let us consider the energy characteristics of the
 waves in chiral
 nihility. The time-averaged Poynting vector can be
 written as
\cite{Landau}:
\begin{equation}
\label{a4}
  \langle\vec P\rangle_t=\frac{1}{4}(\vec E\times\vec H^*+\vec E^*\times\vec
  H)
\end{equation}
 Substituting the fields (\ref{a2}) to (\ref{a4}), we obtain
 \begin{equation}
\label{a5}
  \langle\vec P\rangle_t=e_0^2Y
\end{equation}
where $Y=\eta^{-1}$ is the wave admittance, which is real for
chiral nihility.

Assuming $\E=0,\,\mu=0$ in the expression for the
averaged energy density (\ref{a6}) and substituting the fields (\ref{a2}) into
(\ref{a6}), one obtains
\begin{equation}
\label{a7}
  \langle W\rangle_t=\frac{e_0^2}{2}\left[\omega \frac{\partial\E}{\partial\omega }+
  Y^2\omega \frac{\partial\mu}{\partial\omega }+\frac{2Y}{c}
   \left(\K+\omega \frac{\partial\K}{\partial\omega }\right)\right]
\end{equation}
Since the following holds
\begin{equation}
\frac{\partial(\sqrt{\E\mu})}{\partial\omega }=\frac{\eta}{2}\left(\frac{\partial\E}{\partial\omega }\pm
Y^2\frac{\partial\mu}{\partial\omega }\right)
\end{equation}
we can conclude that the energy velocity
\begin{equation}
\label{a8}
 v_{en}= \frac{\langle P\rangle_t}{\langle W\rangle_t}
\end{equation}
coincides with the group velocity, expressed by equation~(\ref{a3}).

In contrast to the usual chiral medium, where limitations for
material parameters (\ref{limit}) take place and both of
right-hand and left-hand  circularly polarized waves
are forward ones, in chiral nihility the left-hand polarized wave is the
backward wave (here we assumed $\K>0$).

\section{Wave reflection and refraction on a  chiral nihility interface}

Let us consider a linearly polarized plane wave illuminating an
interface
 between vacuum $\E_1=\mu_1=1$ and chiral nihility
$\E_2=\mu_2=0$, $\kappa\neq 0$. In the case of the normal incidence such an incident
wave excites right-hand and left-hand polarized waves in chiral
nihility, having equal
 amplitudes, the same directions of the energy
flow (away from the source), and the opposite
 phase velocities. If the wave impedance in chiral nihility
 is close to that of free space, the reflected wave can be neglected. The
total transmitted electric field can be
 written as
\begin{eqnarray}
 \label{d1}
\vec E & = & \vec E_+ +\vec E_- \nonumber \\
 & = & \frac{e_0}{2}(\vec x_0-{\rm j}\vec
y_0)\exp{(-{\rm j}\K k_0z)}+\frac{e_0}{2}(\vec x_0+{\rm j}\vec
y_0)\exp{({\rm j}\K k_0z)} \nonumber \\
 & = & e_0(\vec x_0\cos{\K k_0 z}-\vec
y_0\sin{\K k_0 z})
\end{eqnarray}

Thus, the total electric field represents the wave of ``standing
phase", whose amplitude changes along $z$-axis in accordance with
(\ref{d1}). Such a wave has infinite phase velocity, but its
 energy
velocity is expressed by equation~(\ref{a8}).
 It can be easily shown that
the proper contributions to the
 Poynting vector from refracted
left-hand and right-hand waves have the same
 signs and their mutual
(interference) time-averaged left-hand--right-hand contribution
 is
equal to zero.  The field forms a standing spiral structure in
space.

\begin{figure}[h]
\centering
\epsfig{file=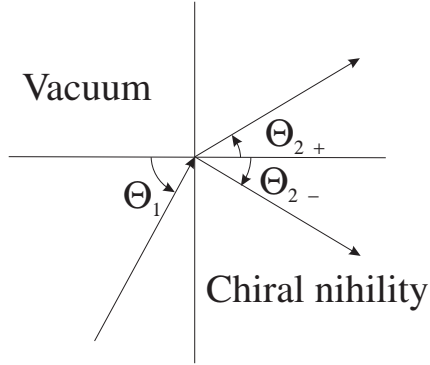, width=6cm}
\caption{An incident plane wave hits the boundary of an isotropic chiral
nihility half space. Note the two refracted rays. The refracted waves are 
circularly polarized, with the opposite sense of rotation. The arrows show the
directions of the power flow.}
\label{Fig1}
\end{figure}

Next, let us consider oblique incidence of a linearly polarized plane
wave from
 vacuum half space (region~1) on an interface with a chiral
nihility half space (region~2) according to Figure~\ref{Fig1}. In the
medium, there will propagate two circularly-polarized  waves.
The Snellius law reads \cite{chibi}
\begin{equation}
\sin{\theta_{2\pm}}=\frac{1}{\pm\K}\sin{\theta_1}
\end{equation}
This means that the two refracted waves propagate at the angles
$\theta_{2\pm},\;\theta_{2-}=-\theta_{2+}$ to the normal.
Anomalous refraction takes place for the left-hand polarized wave, because
that wave is a
backward-wave. We can conclude that in this case
double refraction is possible at an interface
between two isotropic media. Moreover, one of the
rays suffers anomalous refraction, like in the Veselago medium.
One can say that one of the eigenwaves in chiral nihility
sees an equivalent isotropic dielectric, but the other one
sees a backward-wave (Veselago) medium.

\section{Concluding remarks}

The present paper has introduced the concept of chiral nihility
medium. Using the model for a regular lattice of complex particles, a
composite was modeled that displays parameters required for chiral
nihility. The parameters of the material were found to be consistent
with the physical restrictions for complex media.
Analyzing the field equations, we have introduced the concept of
wave impedance in nihility materials defined through a limit
of the permeablity/permittivity ratio. Also, wave propagation
and refraction involving chiral nihility media have been discussed.
It has been found that the eigenwaves are circularly polarized, like
in isotropic chiral media, but one of the eigenwaves is a backward-wave,
like in Veselago media with double negative parameters. Note that materials
with negative parameters are sometimes called ``left-handed" materials
(because the triplet of vectors $\vec E$, $\vec H$, and $\vec k$ is left-handed).
This name can be especially confusing in the present case of chiral nihility,
where the left- or right-hand circularly polarized wave is a backward wave.
As such, handedness of chiral materials with helical inclusions has nothing
in common with the existence of backward waves in Veselago media.
We have found that
an interface between an isotropic chiral nihility material and free space
has a very interesting property of double refraction: the wave is split into
two circularly polarized components, such that one of them is refracted positively,
but the other one is refracted negatively, like in Veselago media.
A helical standing wave pattern is formed in chiral nihility material, if
it is excited by a normally incident plane wave.

\end{document}